\documentclass{PoS}

\usepackage{graphicx}
\usepackage{amsmath}

\title{Observation of Galactic Cosmic Rays and Gamma Rays with the
High Altitude Water Cherenkov Observatory}

\ShortTitle{Gamma Rays Observed by HAWC}

\author{\speaker{Segev BenZvi}\thanks{For the HAWC Collaboration.}\\
        Department of Physics and Astronomy, University of Rochester,
        Rochester, NY, USA\\
        E-mail: \email{sybenzvi@pas.rochester.edu}}

\abstract{The High-Altitude Water Cherenkov Observatory, or HAWC, is carrying
out an unbiased survey of cosmic rays and gamma rays from the Northern
Hemisphere between 100 GeV and 100 TeV. HAWC is currently the only high-uptime
wide-field TeV observatory in operation, and has a robust program to search for
flares and other transient sources of gamma rays. The detector is also well
suited to observe spatially extended regions of gamma-ray emission and
cosmic-ray anisotropy. HAWC recently concluded its first year of data taking
with the complete detector. The results include not only observations of many
known TeV point sources, but also extended emission from Galactic objects like
the Geminga supernova remnant. These results have implications for the origins
of several astrophysical anomalies observed in the cosmic-ray data, such as the
excess of Galactic positrons at Earth. We will describe results from HAWC with
a focus on the observation of cosmic rays and Galactic sources of gamma rays.}

\FullConference{38th International Conference on High Energy Physics\\
		3-10 August 2016\\
		Chicago, USA}

\begin{document}

\section{Introduction}

The High Altitude Water Cherenkov (HAWC) Observatory is a gamma-ray and
cosmic-ray air shower detector located 4100~meters above sea level at
$19^\circ$N latitude in Sierra Negra, Mexico. The detector is operating
continuously with $>90\%$ uptime, including scheduled maintenance periods, and
provides an unbiased wide field-of-view survey of the Northern Hemisphere.

HAWC is sensitive to primary cosmic rays between $100$~GeV and $100$~TeV. At
these energies the detector observes the spectrum of the cosmic rays in a
transition region between satellite-based detectors such as AMS (which observe
primary cosmic rays directly) and ground-based air shower detectors (which
observe primary particles indirectly). HAWC records cosmic-ray showers at an
event rate $>20$~kHz, and the resulting high statistics provide sensitivity to
the anisotropy in the arrival direction distribution of the cosmic rays at the
$10^{-4}$ level \cite{Abeysekara:2014sna}.  HAWC can also observe the strong
deficits (or ``shadows'') in the cosmic ray flux created by the Moon and Sun.
Using the deflection of the shadows in the geomagnetic field, the observatory
can discriminate between positively and negatively charged particles, and hence
can estimate the fluxes of cosmic antiparticles ($\bar{p}$, $e^+$, etc.) at
energies above the limits of direct detectors \cite{BenZvi:2015kga}.

Above $1$~TeV the information recorded by the detector can be used to
effectively filter out cosmic-ray events. Between $1$~TeV and $100$~TeV HAWC is
carrying out an unbiased wide field-of-view survey of gamma rays from the
Northern Hemisphere. The high uptime of HAWC allows for the observation of
transient sources of gamma rays from $2/3$ of the sky each day
\cite{Lauer:2015bza, Wood:2015yma, BenZvi:2015vyo}. Its high energy reach
enables HAWC to distinguish gamma rays originating in $\pi^0$ decay from
inverse Compton scattered photons,
allowing us to search for accelerators of hadronic cosmic rays in the Milky
Way. Finally, thanks to the 2~sr instantaneous field of view of the array, HAWC
is sensitive to gamma rays from very spatially extended galactic and
extragalactic sources \cite{Baughman:2015oga}. These sources include diffuse
emission, which is important for the identification of the origin of cosmic
neutrinos as well as the characterization of gamma rays from Dark Matter
annihilation and decay.

Construction of HAWC ended in March 2015. In this proceeding we present
preliminary results from the first 15 months of data taking with HAWC, and
discuss prospects for the future.

\section{HAWC Operation and Analysis}

\begin{figure}[htbp]
  \includegraphics[width=0.495\textwidth]{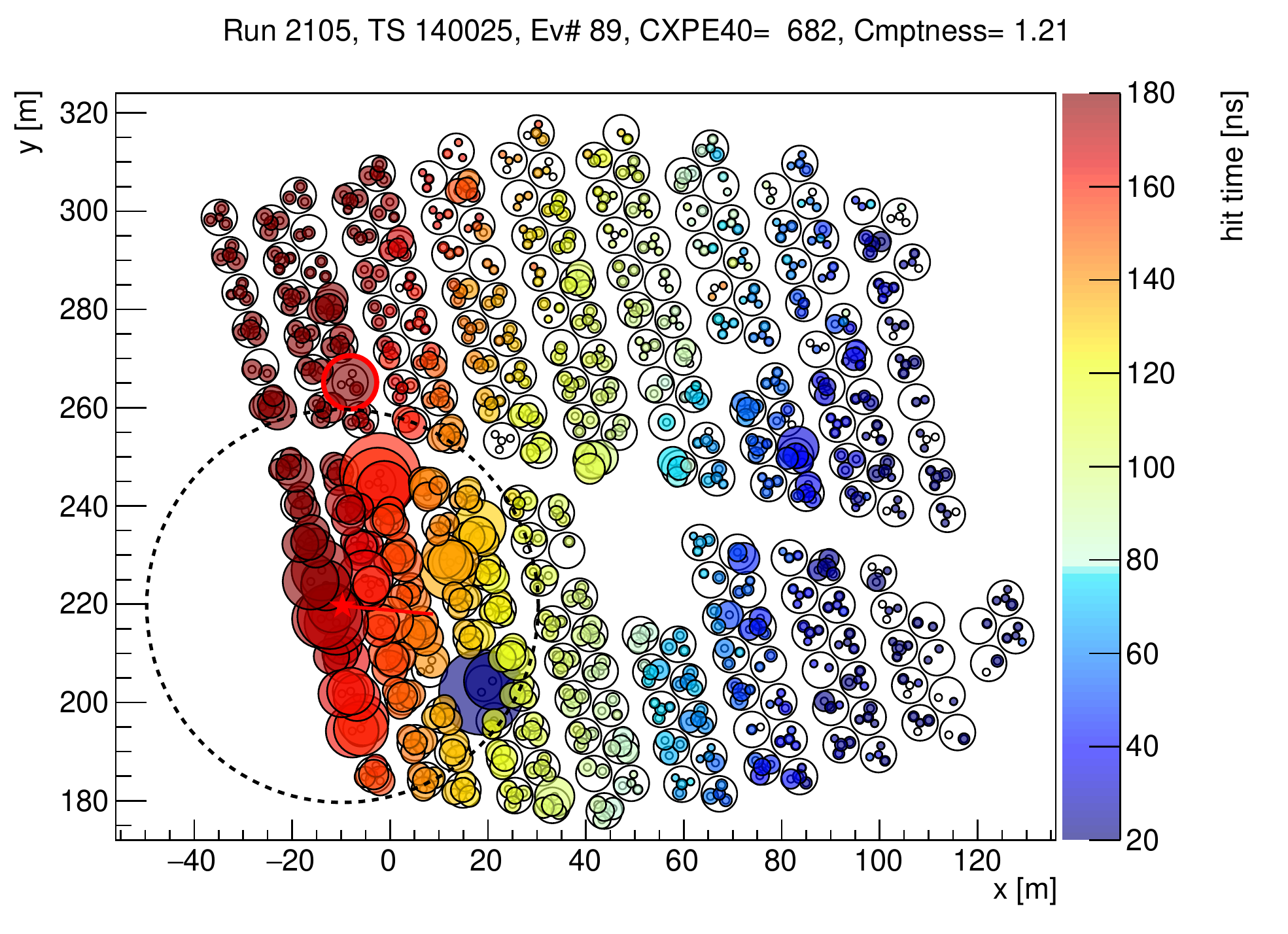}
  \includegraphics[width=0.495\textwidth]{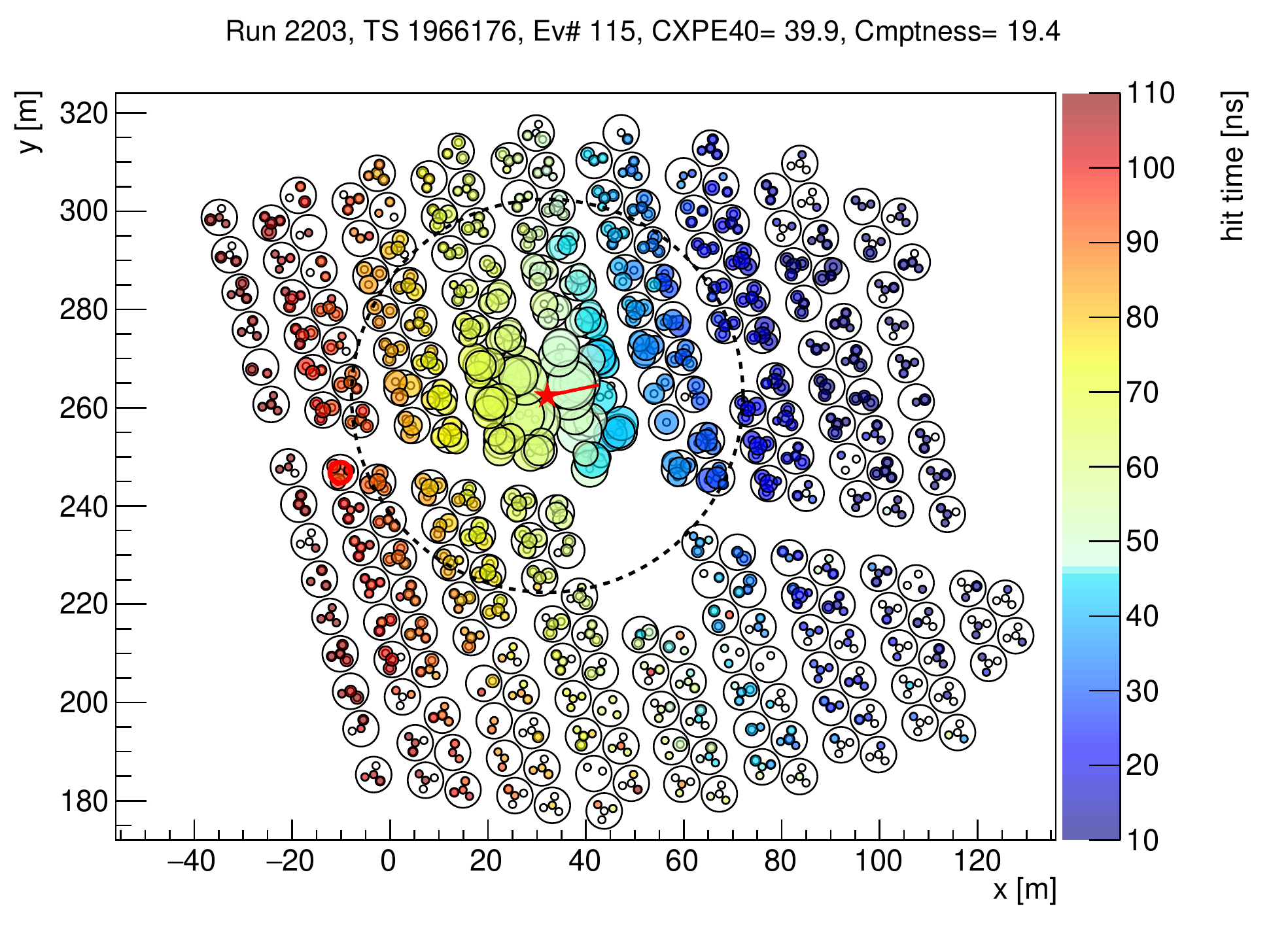}
  \caption{\label{fig:ghsep} {\sl Left}: event display showing a cosmic-ray
  shower triggering the complete HAWC array. The colors indicate the trigger
  times of PMTs in the detector, and the circle area is proportional to the
  number of photoelectrons recorded by each PMT. Small empty circles indicate
  untriggered PMTs. {\sl Right}: a gamma-ray event.}
\end{figure}

The HAWC Observatory comprises 300 water Cherenkov detectors (WCDs) covering an
area of 20,000~m$^2$ \cite{Smith:2015wva}. The layout of the WCDs is
illustrated in Fig.~\ref{fig:ghsep}. Each WCD is a cylindrical 5~m $\times$
7.3~m steel tank.  Its interior is lined with a plastic bladder and filled with
200,000~L of purified water. Four photomultipliers anchored to the bottom of
the WCD record the Cherenkov light produced when charged particles in an air
shower pass through the detector.

The relative timing of charged particle hits in the WCDs is used to reconstruct
the arrival direction of the primary particle which initiated the air shower,
and the number of triggered channels in the detector is roughly proportional to
the energy of the primary cosmic ray or gamma ray. The accuracy of the
directional reconstruction ranges from about $1^\circ$ below $1$~TeV to
$0.1^\circ$ above $10$~TeV.

As shown in Fig.~\ref{fig:ghsep}, cosmic-ray showers are characterized by
``clumpy'' deposits of charge at large distances from the core of the air
shower, while gamma rays appear to have a relatively low variance in the charge
as a function of distance from the shower core. Cuts on the spatial
distribution of charge recorded in each trigger are used to filter out
cosmic-ray events and keep gamma-ray showers. The efficiency of the cosmic-ray
cuts improves as a function of energy, ranging from about 5\% at $1$~TeV to
0.1\% at 10~TeV, while the gamma ray passing efficiency is about 75\%.

The search for gamma-ray sources is carried out using a binned analysis of fine
spatial pixels and coarse shower size bins using the fraction of active PMTs
triggered. The cosmic-ray background filter and detector point spread function
are optimized in each shower size bin.  Spatial and spectral models of gamma-ray
sources are then forward-folded through the detector response using simulated
events and fit to data in the shower size bins. The fit is performed using a
maximum likelihood
\[
  \ln{\mathcal{L}}(\vec{n}|\vec{\theta}) =
    \sum_{i=1}^{N_\text{bin}}
    \sum_{j=1}^{N_\text{pix}}
    n_{ij}\ln{\lambda_{ij}}(\vec{\theta}) - \lambda_{ij}(\vec{\theta}) - \ln{n_{ij}!}
\]
%
%
where the observed counts $n$ are compared to the model counts $\lambda$, which
are the sum of background and signal contributions: $\lambda_k=B_k+\sum_l
f_{kl}(\vec{\theta})$. In principle several sources can contribute to the model
counts in each pixel, and the source spatial and spectral parameters
$\vec{\theta}$ are estimated by maximizing $\ln{\mathcal{L}}$. To estimate the
significance of a detection, we use the test statistic defined by the
likelihood ratio of signal plus background versus background alone:
\[
  \text{TS} = 2\Delta\ln{\mathcal{L}}.
\]
The significance of the detection is given by $\sqrt{\text{TS}}$.

\section{Survey of the Inner Galaxy}

\begin{figure}[htbp]
  \includegraphics[width=\textwidth]{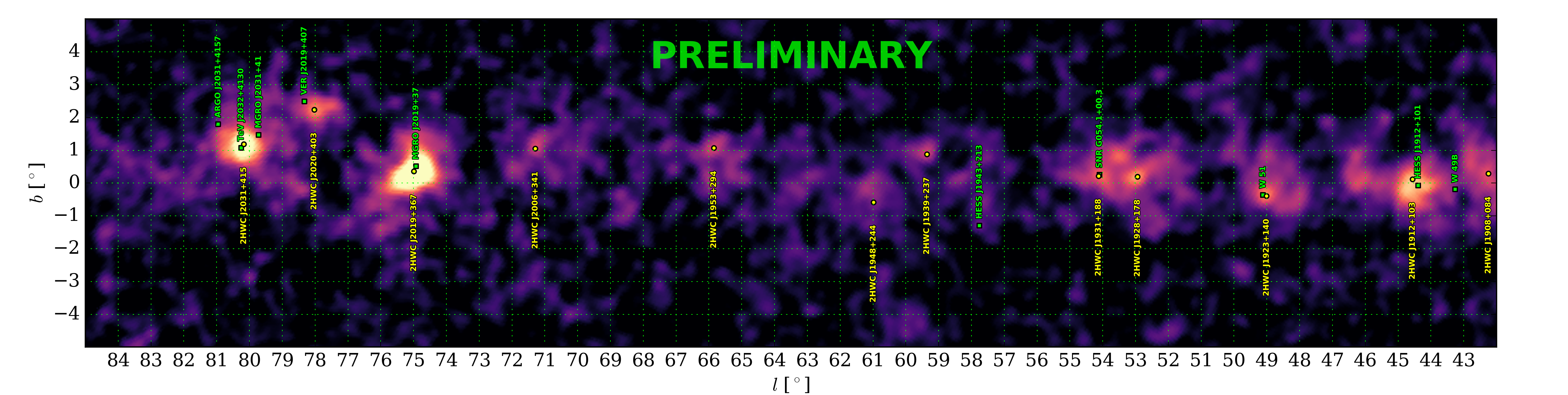}
  \includegraphics[width=\textwidth]{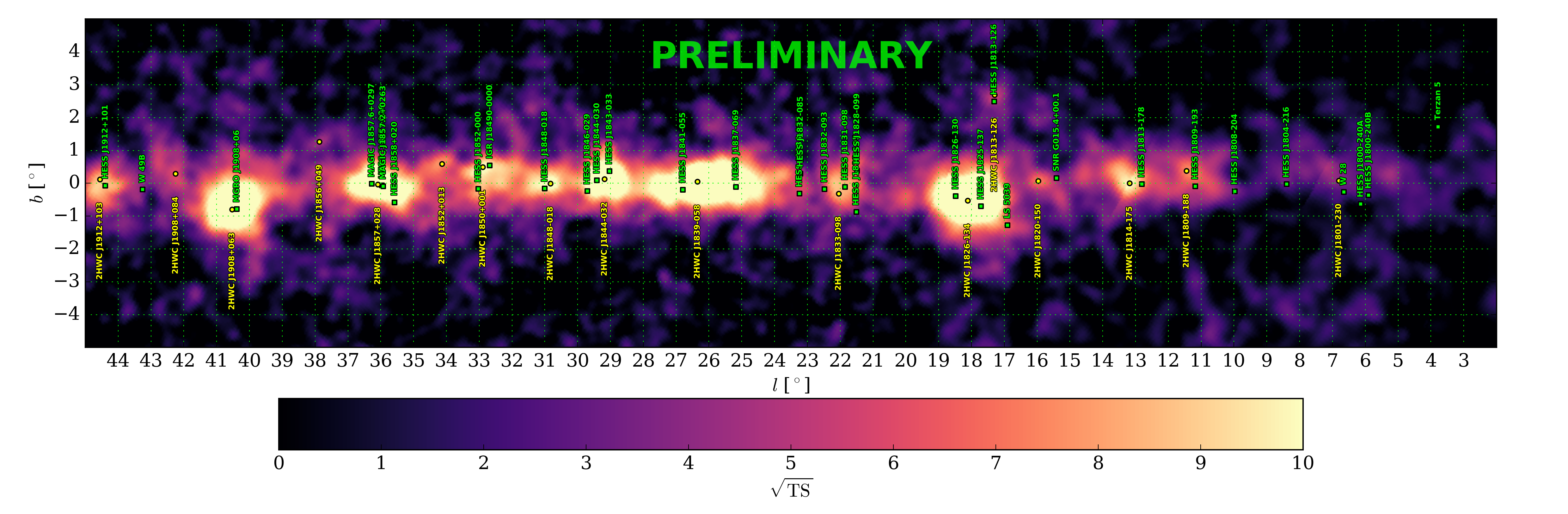}
  \caption{\label{fig:innergal} Sources of gamma rays between $1$ and $100$~TeV
  along the region of the Galactic Plane observed by HAWC.}
\end{figure}

Figure~\ref{fig:innergal} shows the region of the inner Galaxy visible to HAWC
from its location in central Mexico. The skymap shows $\sqrt{\text{TS}}$ as a
function of galactic longitude going from the Galactic Center region (bottom
right) to the Cygnus region (top left), one of the largest stellar nurseries
in the Milky Way.
The green labels in the skymap indicate sources discovered by previous TeV
surveys in the Northern and Southern Hemispheres, while the yellow labels
indicate local maxima in $\sqrt{\text{TS}}$ that were identified as gamma-ray
source candidates in 15 months of HAWC data. Due to the statistical penalty
incurred by a blind search of this region, $\sqrt{\text{TS}}\approx7$
corresponds to a $5\sigma$ detection after trials.

\begin{figure}[htbp]
  \centering
  \includegraphics[width=0.5\textwidth]{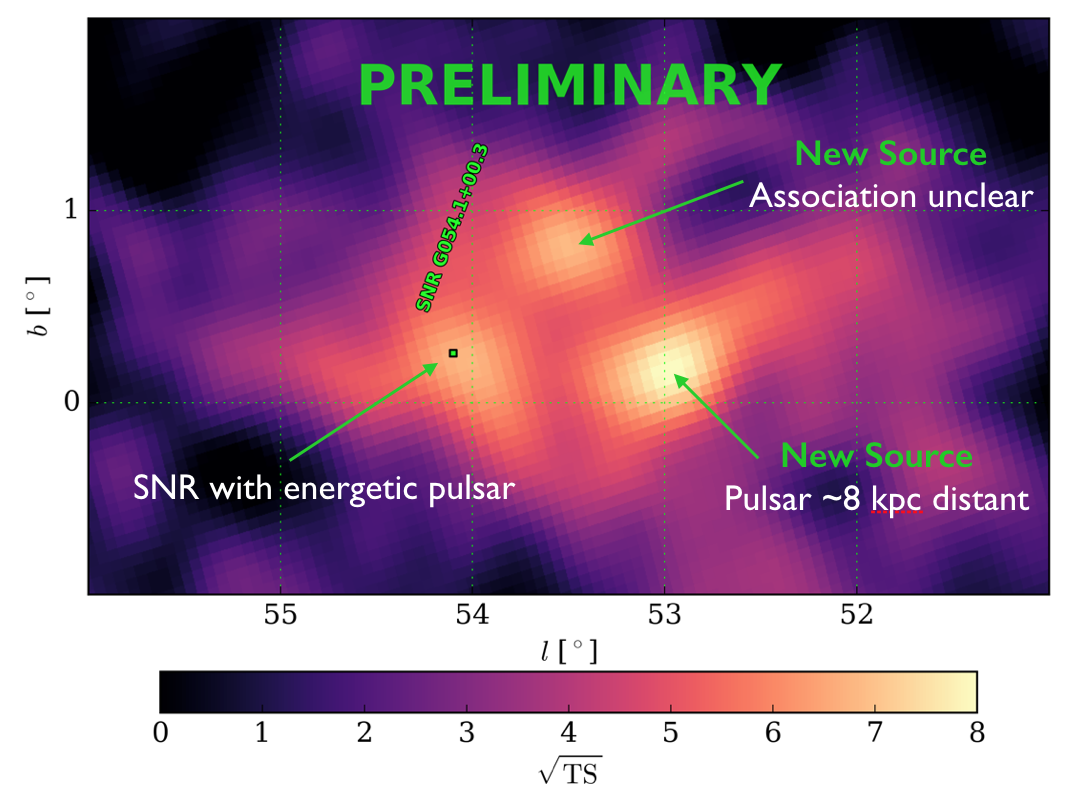}
  \caption{\label{fig:executioner} Zoom-in of a region of the inner Galaxy
  showing two previously unknown point sources of gamma rays.}
\end{figure}

Several new source candidates with no TeV counterparts have been observed in
this data set, and these discoveries are currently being followed up at other
wavelengths. One such region of interest is shown in
Fig.~\ref{fig:executioner}. The region contains one previously known TeV
source, a nearby supernova remnant with an energetic pulsar, and two new
sources discovered by HAWC. The detection of the new sources in a region
previously explored by TeV detectors exhibits the power of the HAWC
Observatory; due to its wide field of view it can cover many regions of the sky
without dedicated observing time. In addition, while HAWC is not as sensitive
to point sources of gamma rays as other TeV observatories, it is less affected
by systematic uncertainties in the sky background that can prevent the
detection of point sources.

\section{Very Extended Emission: Geminga and PSR B0656+14}

\begin{figure}[htbp]
  \includegraphics[width=0.495\textwidth]{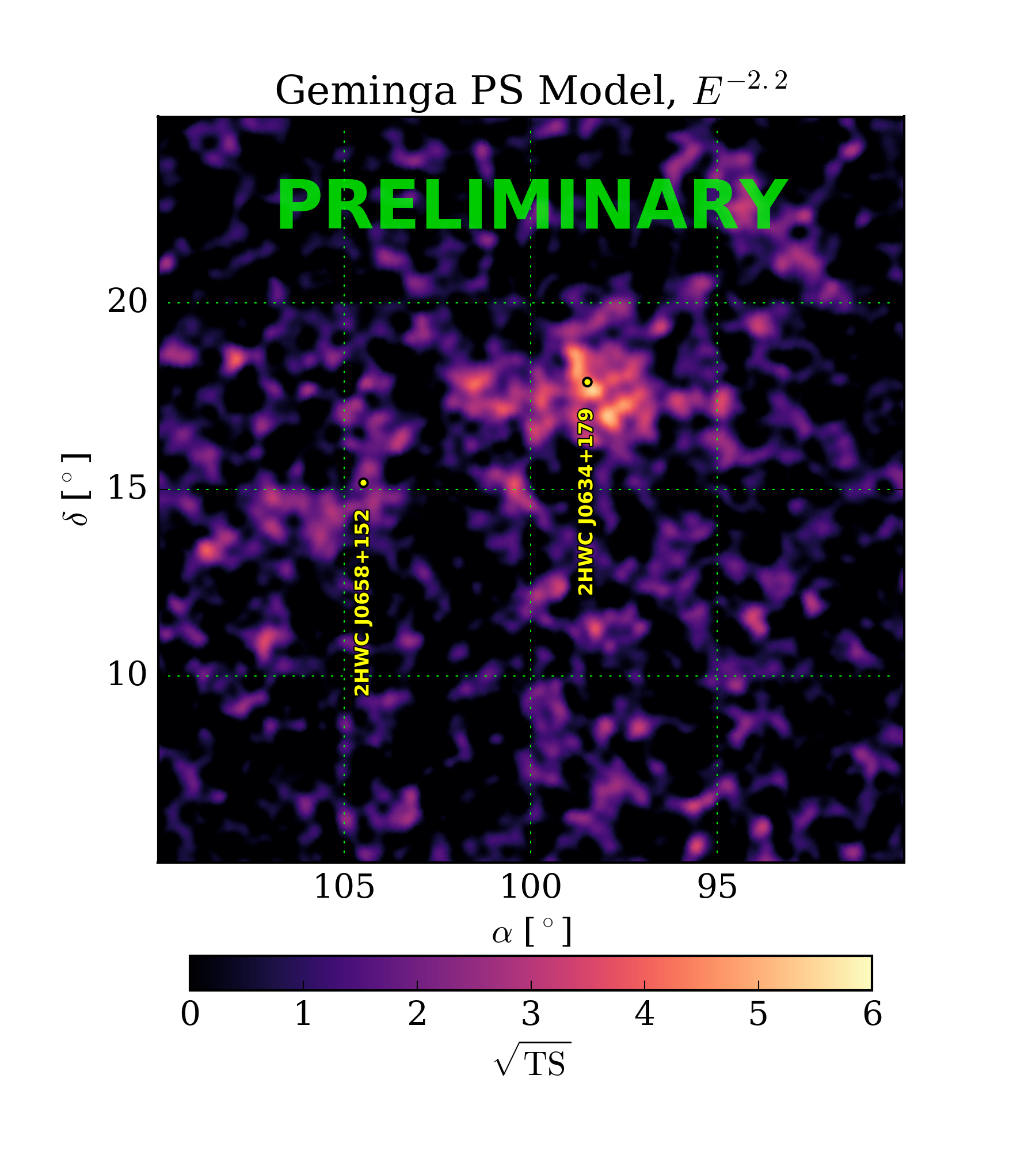}
  \includegraphics[width=0.495\textwidth]{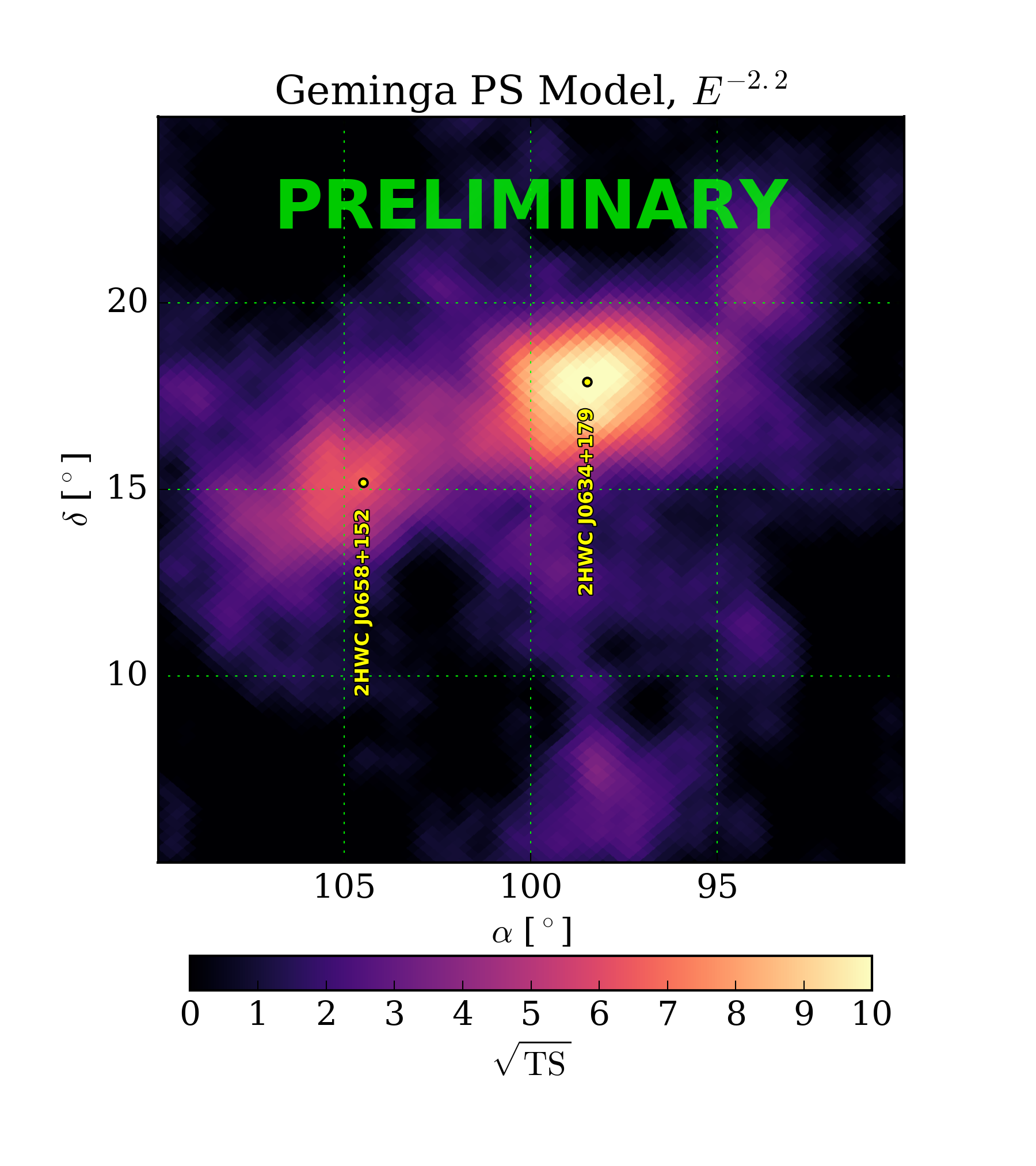}
  \caption{\label{fig:geminga} Spatially extended regions of emission from
  Geminga (2HWC~J0634+179) and PSR B0656+14 (2HWC~J0658+152), shown in
  equatorial coordinates. The left plot shows the region after smoothing by the
  detector point spread function; the right plot shows the same region smoothed
  by a $2^\circ$ top hat function.}
\end{figure}

Among the new sources of gamma rays detected by HAWC are two very spatially
extended regions of emission in the outer Galaxy: Geminga and PSR B0656+14.
Both objects are pulsar wind nebulae located $\sim250$~pc from Earth. Neither
object has been observed previously at TeV, though the Milagro Collaboration
reported a sub-$5\sigma$ detection of extended emission from Geminga in 2009
\cite{Abdo:2009ku}. While the Geminga pulsar has been observed in X-rays and
GeV gamma rays \cite{Caraveo:2003, Abdo:2010wp}, the observations with HAWC are
the first time extended emission has been observed from its nebula. The lack of
multi-wavelength observations of the nebula is not currently understood.

Due to their proximity to the Solar system, Geminga and PSR B0656+14 are strong
candidates for the source of the local electron and positron flux observed at
Earth. These objects may also be the source of the local excess of positrons
above the rate expected from cosmic ray interactions in the interstellar
medium. Since the positron excess is also a signature of dark matter
annihilation, the association of the excess with nearby particle accelerators
is of considerable interest. We are currently using the observations with HAWC
to investigate this possibility.

\section{Conclusions}

HAWC has been operating continuously since March 2015 and has accumulated one
of the world's largest data sets of cosmic rays and gamma rays above $1$~TeV.
Among the first observations are the detection of $>30$ point sources in the
inner Galaxy, which includes 9 objects not previously detected at TeV. Very
extended emission has also been observed from the nebulae surrounding Geminga
and PSR B0656+14, and we are exploring the possibility that these nearby
objects contribute significantly to the local $e^+e^-$ flux at Earth.

HAWC is currently funded to operate through 2020.  The HAWC collaboration is
now constructing a high-energy ``outrigger'' extension of the central array,
which will increase the effective area to gamma rays above $10$~TeV by up to a
factor of four. We expect the high-energy extension to be in operation in 2017.

%
%

\section*{Acknowledgments}
%
This work was supported by the Department of Energy Office of High Energy
Physics under grant number DE-SC0008475.

\bibliographystyle{JHEP}

\bibliography{ichep2016}

\end{document}